\documentclass{article}

\usepackage{PRIMEarxiv}

\usepackage[utf8]{inputenc} % allow utf-8 input
\usepackage[T1]{fontenc}    % use 8-bit T1 fonts
\usepackage{hyperref}       % hyperlinks
\usepackage{url}            % simple URL typesetting
\usepackage{booktabs}       % professional-quality tables
\usepackage{amsfonts}       % blackboard math symbols
\usepackage{nicefrac}       % compact symbols for 1/2, etc.
\usepackage{microtype}      % microtypography
\usepackage{lipsum}
\usepackage{fancyhdr}       % header
\usepackage{graphicx}       % graphics
\graphicspath{{media/}}     % organize your images and other figures under media/ folder
\usepackage{cite}
\usepackage{epsfig}
\usepackage{graphics}
\usepackage{amsmath} 
\usepackage{color,soul}

\title{Autonomous Wheel Loader Trajectory Tracking Control Using LPV-MPC
}

\author{
  Ruitao Song, Zhixian Ye, Liyang Wang\\
  Baidu RAL\\Sunnyvale, CA, 94089, USA. \\
  \texttt{\{ruitaosong, zhixianye, liyangwang\}@baidu.com} \\
   \And
  Tianyi He 
  \thanks{\textit{\underline{Corresponding author}}} \\
  Utah State University \\
  Logan, Utah, 84342, USA.\\
  \texttt{tianyi.he@usu.edu} \\
  \AND
  Liangjun Zhang \\
  Baidu RAL \\
  Sunnyvale, CA, 94089, USA.\\
  \texttt{liangjunzhang@baidu.com} \\
}

\begin{document}
\maketitle

\begin{abstract}
In this paper, we present a systematic approach for high-performance and efficient trajectory tracking control of autonomous wheel loaders. With the nonlinear dynamic model of a wheel loader, nonlinear model predictive control (MPC) is used in offline trajectory planning to obtain a high-performance state-control trajectory while satisfying the state and control constraints. In tracking control, the nonlinear model is embedded into a Linear Parameter Varying (LPV) model and the LPV-MPC strategy is used to achieve fast online computation and good tracking performance. To demonstrate the effectiveness and the advantages of the LPV-MPC, we test and compare three model predictive control strategies in the high-fidelity simulation environment. With the planned trajectory, three tracking control strategies LPV-MPC, nonlinear MPC, and LTI-MPC are simulated and compared in the perspectives of computational burden and tracking performance. The LPV-MPC can achieve better performance than conventional LTI-MPC because more accurate nominal system dynamics are captured in the LPV model. In addition, LPV-MPC achieves slightly worse tracking performance but tremendously improved computational efficiency than nonlinear MPC. A video with loading cycles completed by our autonomous wheel loader in the simulation environment can be found here: \url{https://youtu.be/QbNfS_wZKKA}.
\end{abstract}

% keywords can be removed

\section{Introduction}
\begin{figure}[ht]
\centering
\includegraphics[width = 0.55\linewidth]{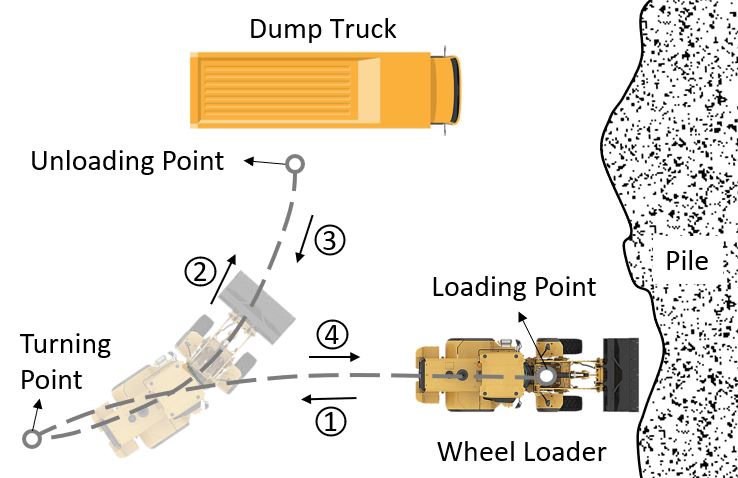}
\caption{A typical loading cycle of the wheel loader.}
\label{fig:loadingcycle}
\end{figure}

Wheel loaders are often used to transport materials on mining and construction sites, as shown in Fig.~\ref{fig:loadingcycle}. Currently, wheel loaders are mostly controlled by trained human operators. The prolonged training process leads to global labor shortages for operating heavy mining and construction equipment. Besides life-threatening incidents, human operators often have to operate the wheel loaders in extreme working conditions, such as heavy dust and extreme temperatures \cite{zhang2021autonomous}. Moreover, the frequent acceleration, deceleration, and steering actions of wheel loaders pose considerable challenges to the wheel loader drivers, making it impossible to maintain high working efficiency and quality over long operation periods \cite{shi2020planning}. These issues stimulate the critical demands of autonomous systems equipped on wheel loaders and other articulated vehicles in extreme working conditions \cite{hemami2009overview}. 

Wheel loaders consist of a front and rear body connected by a hinge joint and a swing ring. In the articulated steering process, the front and rear vehicle bodies are connected by the hydraulic actuators to steer the vehicle \cite{oh2016gear}. This mechanism reduces the turning radius and improves the maneuverability of the vehicle, which gives it good adaptability in various operating environments \cite{frank2018optimal}. However, this steering mechanism introduces highly nonlinear dynamics, hence imposes extra complexity on the trajectory planning and tracking control problem. 

Alshaer et al. \cite{alshaer2013path} improved the Reeds and Shepp path planning method to an autonomous wheel loader and applied PID controller for trajectory tracking. Choi et. al. \cite{choi2015constrained} adopted the $A^\ast$ algorithm to optimize the global path of an articulated vehicle. Shi et. al. \cite{shi2020planning} developed an adaptive MPC controller to track the trajectory generated by an algorithm based on rapidly exploring random tree. These trajectory planning methods did not consider the full kinematic constraints of the wheel loader, and the articulated angle was not considered during planning. Therefore, the planned trajectory is hard for the downstream controller to follow, and the planning algorithm usually has to consider extra safety distance between the vehicle and obstacles. Shi et. al.\cite{shi2020planning} considered the path curvature in the adaptive MPC controller to improve the tracking performance. Nayl et al. \cite{nayl2013modeling} proposed a bug-like path planning technique and used a switching MPC controller considering varying slip angles and different velocities. However, these MPC controllers used the linearized system model at each time instant as the constant predictive model during the prediction horizon. This will lead to a compromised performance compared with nonlinear MPC controllers because of the loss of nonlinear dynamics in the prediction horizon. In \cite{nayl2018design}, a sliding mode controller was proposed, but the tracking performance of the articulated angle was not considered by the controller. Similar issue can also be found in \cite{li2020research} and \cite{tian2021active}. 

The Linear Parameter Varying (LPV) control has received lots of attention from academia and industry to address the nonlinear systems, including automotive~\cite{zhang2014lpv,mohammadpour2012control}, aerospace~\cite{he2018application}, medical engineering~\cite{colmegna2015switched} and robotics~\cite{san2021disturbance}. There are emerging applications using LPV model in model predictive control on autonomous driving. Alcala et. al. \cite{alcala2020autonomous} developed a trajectory tracking controller using LPV-MPC for the race car. Cheng \cite{cheng2020model} used the model predictive control on LPV model of lateral dyanmics to design steering actions in path tracking. The LPV model has a linear representation of system matrices, which are dependent on scheduling parameters embedded from nonlinear dynamics. In this way, the control design can take the benefit of linear formulation and address the nonlinear dynamics. The computational complexity of the optimization problem can be greatly simplified from nonlinear programming.  Since the scheduling parameters still involve nonlinear dynamics of state and control inputs, the system dynamics don't lose nonlinearity. Therefore, the LPV model will produce more accurate predictions than adaptive LTI-MPC, whose predictive model is obtained by iterative linearization along the planned trajectory.  

In this paper, we present a systematic approach for autonomous wheel loader driving system, which consists of offline high-performance planning by nonlinear MPC and online tracking control using LPV-MPC. With the nonlinear model and refined constraints on states and control inputs, a high-performance trajectory is planned using nonlinear MPC containing information of all scheduling parameters used in the LPV model. Due to the heavy computational complexity, trajectory planning is conducted offline to generate the nominal trajectory, and the relatively static environment factors and constraints are considered. After planning is done, the LPV-MPC strategy is used to track the nominal trajectory. With the sequences of state-control trajectory, the nonlinear model is converted into a quasi-LPV model by writing the nonlinear terms to time-varying scheduling parameters. The LPV model is then used in MPC as the predictive model to capture the nominal nonlinear dynamics in the prediction horizon.  

The main contributions of this work are three-fold: 1) An architecture consisting of nonlinear MPC for offline planning and LPV-MPC for online tracking control, that is able to complete the whole loading cycle of the wheel loader; 2) Developing the LPV model from the nonlinear model for wheel loader tracking control; 3) Demonstrating the outstanding tracking performance of LPV-MPC but with great computational efficiency. Besides the heavy-duty wheel loaders, the proposed architecture of planning and control can be easily applied to other articulated vehicles in an unstructured environment. 

The rest of this paper is organized as follows. Section~\ref{model_planning} introduces the nonlinear model of the wheel loader and the nonlinear MPC technique to plan the state-control trajectory. Section~\ref{tracking_control} formulates the LPV-MPC problem for tracking the planned trajectory. Section~\ref{simulation} then presents the simulations in a high-fidelity environment and compares the tracking performances and computational burden of LPV-MPC with nonlinear MPC and adaptive LTI-MPC. At last, conclusions are made and future work is discussed. 
\section{Modeling and Trajectory Planning of Articulated vehicle}\label{model_planning}
\subsection{Nonlinear model}\label{nl_model}
The wheel loader considered in this paper is mainly operated in low-speed conditions, in which the slip effect is minor. Therefore, the tire slip angle and lateral forces acting on the tires are not considered. The kinematic model is used for both trajectory planning and tracking control. 

\begin{figure}[ht]
\centering
\includegraphics[width = 0.42\linewidth]{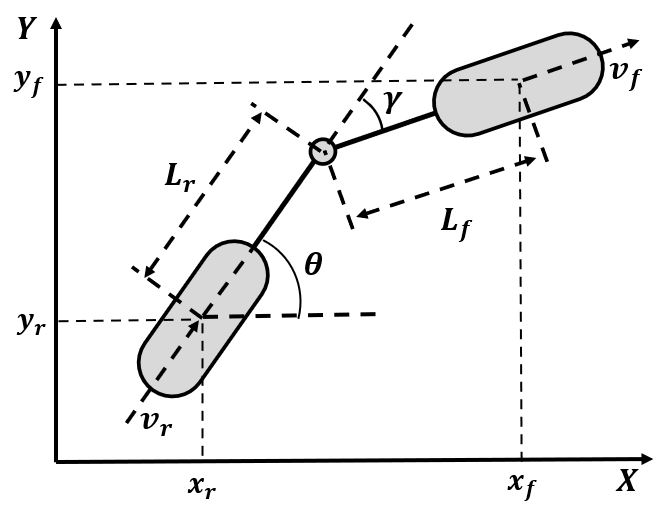}
\caption{The wheel loader model used for planning and control design.}
\label{fig:wl_model}
\end{figure}

The bi-cycle kinematic of the wheel loader is depicted in Fig.~\ref{fig:wl_model}, where each axle being composed of two wheels is replaced by a unique wheel. $\left(x_f,y_f\right)$ and $\left(x_r,y_r\right)$ are the positions of the center points of the front and rear wheel axle, respectively; $\theta$ is the rear vehicle body heading angle; $\gamma$ is the articulated angle, that is also the difference between the front and rear body heading angles; $L_f$ and $L_r$ represent the distance from the articulation point to the front and rear axle, respectively. $v_f$ and $v_r$ are the speeds of the front and rear vehicle body. 

It is assumed that the vehicle's speed and acceleration are small and the slip effect is neglectable. In other words, the direction of the front (rear) body speed is aligned with the heading of the front (rear) body. According to~\cite{nayl2013modeling}, the kinematics of the wheel loader can be described by the following equations: 
\begin{equation}
    \left\{
    \begin{aligned}
     & \dot{x}_f = v_f cos\left(\theta + \gamma\right)~or~\dot{x}_r = v_r cos\theta\\
     & \dot{y}_f = v_f sin\left(\theta + \gamma\right)~or~\dot{y}_r = v_r sin\theta\\
     & \dot{\theta} = \frac{v_f sin\gamma - L_r \dot{\gamma} cos\gamma}{L_f cos\gamma + L_r}
    \end{aligned}
    \right.
\label{eqn:ss_con}
\end{equation}

Once the rear body heading angle and the articulated angle are known, the state of the wheel loader can be fully described by the position of either the front or rear vehicle body. Therefore, the state and control vectors can be denoted as:
\begin{equation}
    x=\begin{bmatrix}x_r \\ y_r \\ \theta \\ \gamma \end{bmatrix}~or~\begin{bmatrix}x_f \\ y_f \\ \theta \\ \gamma \end{bmatrix},~ u = \begin{bmatrix} v \\ \dot{\gamma} \end{bmatrix}
\label{eqn:ss_vec}
\end{equation}
where the state vector has two representations: one using front body position and the other using rear body position. Note that, the front body position can be calculated from the rear body position and their relative angle with the length of front and rear bodies, therefore the whole-body dynamics can be fully represented by either of the state vectors. To facilitate calculation, the state representation is selected based on the moving direction of the wheel loader, which will be explained in Section~\ref{tracking_control}. With states and control inputs in~\eqref{eqn:ss_vec}, nonlinear dynamics~\eqref{eqn:ss_con} can be rewritten as $\dot{x} = f(x,u)$. The discrete nonlinear state space representation of the system can be derived using the Euler method:
\begin{equation}
    x(k+1) = x(k) + T_{s} f(x(k),u(k)),
\label{eqn:ss_dis}
\end{equation}
where $T_{s}$ is the step size, and $k$ denotes the step index. 

\subsection{Planning using nonlinear MPC}

Different from the passenger vehicles driving on the highway, this paper addresses the case that the wheel loader is operated in an open and unstructured environment, see Fig.~\ref{fig:loadingcycle}. In the task of excavating and loading, the wheel loader performs a Y-shaped curve between the loading and unloading sites \cite{dadhich2016key}. As shown in Fig.~\ref{fig:loadingcycle}, the loading cycle can be decomposed into four steps in the trajectory planning: 1) load material and retract from the pile; 2) approach to the dump truck and unload material; 3) retract from the truck; 4) approach to the pile. 

In this paper, the trajectory planner firstly generates the Y-shaped trajectory from the loading point to the dump truck (step 1 and 2), then calculates another trajectory from the truck to the loading point (step 3 and 4) again to complete the whole loading cycle. At the loading and unloading points, the desired position, heading angle, and articulated angle are determined by the pile and truck locations. In other words, the initial and desired final states of the wheel loader are fixed. The trajectory planner needs to calculate a feasible trajectory without collision considering the kinematics of the wheel loader. The trajectory planning problem at time instance $k$ can be formulated into a nonlinear optimization problem:
\begin{equation}\label{eqn:cost_plan}
    \begin{aligned}
    \min\limits_{U}~&{J(x(k),U)} = \sum_{i=0}^{N-1}\left( {||R~u(i|k)||}_{2} + {||R_d~ \Delta u(i|k)||_2}\right)\\
    s.t.~& x(i+1|k) = x(i|k) + f\left(x(i|k),u(i|k)\right)T_{s},\\
    & x(0|k) = x_0,~x(N|k) = x_f,~x(i|k) \in \mathbf{X},\\
    & u(0|k) = u_0,~u(N|k) = u_f,~U \in \mathbf{U},\\
    & D^2_{sf}-D^2(x(i|k)) < 0,\\
    & -\gamma_{max} < \gamma < \gamma_{max},\\
    & U = col\left\{u(0|k), \dots, u(N-1|k) \right\},
    \end{aligned}
\end{equation}
where $N$ is the prediction horizon. $x(i|k)$ and $u(i|k)$ denote  the predicted values of the model state and input, respectively, at time $k + i$ based on the information that is available at time $k$. $\Delta u(i|k) = u(i|k)-u(i-1|k)$ is included in the cost function to smooth the trajectory. Since the problem is formulated to calculate the trajectory from the initial state to the target state, $x(i|k)$ is not considered in the cost function. The initial and target states and desired control actions ($x_0, x_f, u_0, u_f$) are considered as equality constraints. $\gamma_{max}$ is the maximum allowed articulated angle. $D^2(x(i|k))$ is the distance to the two obstacles: the pile and the dump truck (see Fig.~\ref{fig:loadingcycle}), and $D^2_{sf}$ is the safety distance. The nonlinear MPC problem finds the optimal control sequence ($U$) during the given time horizon defined by $N$ and $T_{s}$. The generated trajectory contains the desired vehicle state at every time step along with the associated control command.

In this paper, the trajectory planner is not executed at every time instant due to limited computational capacity. Since the application scenario does not consider moving obstacles, the trajectories are calculated offline with initial and target states corresponding to the loading and unloading points. The location of the turning point (see Fig.~\ref{fig:loadingcycle}) is determined directly by the nonlinear MPC planner.

\section{Tracking Control using LPV-MPC}\label{tracking_control}

The LPV-MPC strategy has advantages over conventional nonlinear MPC and adaptive LTI-MPC. Firstly, the computational complexity is greatly reduced from nonlinear MPC. By embedding the nonlinear system along the planned trajectory to an LPV model, the MPC optimization problem renders a QP problem with linear time-varying system matrices. Secondly, LPV-MPC outperforms adaptive LTI-MPC in tracking performance when addressing nonlinear dynamics. For adaptive LTI-MPC, the linearized model at the current operating point is used to represent the system dynamics in the prediction horizon. On the contrary, the LPV model embeds the nonlinear dynamics into scheduling parameters, hence the Jacobian matrices along the nominal trajectory are obtained to represent the nominal linear time-varying dynamics. In the LPV-MPC tracking control, the state deviations (tracking error) from nominal trajectory are constrained in a small bound. At the bounded region close to the nominal state-control trajectory, the LPV model can capture the nonlinear dynamics evolving at each step in the prediction horizon. 

\subsection{LPV model embedded from nonlinear model}
The nonlinear state space equation represented in~\eqref{eqn:ss_dis} can be linearized around each point along the trajectory, $({x}^{\ast}(k), {u}^{\ast}(k))$, calculated by the planner. Using first-order Taylor expansion,~\eqref{eqn:ss_dis} can be approximated as:
\begin{equation}
\begin{aligned}
    x(k+1) = &x(k) + T_{s}\Big[f({x}^{\ast}(k),{u}^{\ast}(k)) \Big.\\
    &+ \frac{\partial f({x}^{\ast}(k),{u}^{\ast}(k))}{\partial x} (x(k)-{x}^{\ast}(k))\\
    &\Big.+ \frac{\partial f({x}^{\ast}(k),{u}^{\ast}(k))}{\partial u} (u(k)-{u}^{\ast}(k))\Big].
\end{aligned}
\end{equation}
Let $x_{e}(k):=x(k)-{x}^{\ast}(k)$ denote the tracking error at each time instant and $u_{e}(k):=u(k)-{u}^{\ast}(k)$ denote the tracking control inputs, the above equation can be written as:
\begin{equation}
\begin{aligned}
    x_{e}(k+1) = &\left[I + T_{s}\frac{\partial f({x}^{\ast}(k),{u}^{\ast}(k))}{\partial x} \right]x_{e}(k)
    + T_{s}\frac{\partial f({x}^{\ast}(k),{u}^{\ast}(k))}{\partial u} u_{e}(k).
\end{aligned}
\end{equation}
The Jacobians can be derived with nonlinear terms 
\begin{equation}
    \frac{\partial f}{\partial x} = \left[ \begin{array}{cccc}
        0  & 0 & -v_{f} sin(\theta) & 0 \\
        0 & 0 & v_{f} cos(\theta) & 0\\ 
        0 & 0 & 0 & f_{4} \\
        0 & 0& 0& 0
    \end{array}       \right],
\end{equation}
where $f_{4} = \frac{v cos\gamma + L_{r}\dot{\gamma}sin\gamma}{L_{f}cos\gamma + L_{r}} + \frac{(L_{f}sin\gamma)(v sin\gamma - L_{r} \dot{\gamma}cos\gamma)}{(L_{f}cos\gamma + L_{r})^{2}}$,
\begin{equation}
    \frac{\partial f}{\partial u} = \left[ \begin{array}{cc}
    cos(\theta) & 0 \\
    sin(\theta) & 0 \\
    \frac{sin\gamma}{L_{f}cos\gamma + L_{r}} & \frac{-L_{r}cos\gamma}{L_{f}cos\gamma + L_{r}} \\
    0 & 1
    \end{array}       \right].
\end{equation}
With the nominal trajectory of state $x_{f},y_{f}, \theta, \gamma$ and control inputs $v_{f}, \dot{\gamma}$, the Jacobian matrix in the prediction horizon can be easily calculated by plugging in states and control values. The nonlinear functions in Jacobian matrices are embedded into scheduling parameters, and the discrete-time LPV model is thus derived as:
\begin{equation}
    \begin{array}{ll}
    x_{e}(k+1) & = A({x}^{\ast}(k),{u}^{\ast}(k))~x_{e}(k)
                + B({x}^{\ast}(k),{u}^{\ast}(k)) ~u_{e}(k) 
    \end{array}
\end{equation}
where, 
$A({x}^{\ast}(k),{u}^{\ast}(k)) = I + T_{s}\frac{\partial f}{\partial x}|_{{x}^{\ast}(k),{u}^{\ast}(k)}$,   $B({x}^{\ast}(k),{u}^{\ast}(k)) =  T_{s}\frac{\partial f}{\partial u}|_{{x}^{\ast}(k),{u}^{\ast}(k)} $.

In the prediction horizon, a sequence of $A_{k}, B_{k}$ can be computed from the planned trajectory, and then the MPC problem will be formulated to solve the optimal tracking control action to follow the planned trajectory. For ease of expression, we simply write scheduling parameters of nonlinear functions in the Jacobian matrix as $\rho(k)$. 

\subsection{LPV-MPC problem formulation}

The fundamental idea of LPV-MPC is that, by previewing the scheduling parameter sequences, the optimal control sequences to track the nominal state-control trajectory in a finite horizon can be optimized to minimize the tracking error. The optimization is conducted repetitively in the receding horizon, and only the first action at each iteration is considered in the output optimal sequence. At each time instance $k$, the problem of LPV-MPC is expressed in~\eqref{LPV_MPP}:

\begin{equation}\label{LPV_MPP}
	\begin{aligned}
		\min\limits_{U_{e}}~ &{J(x_{e}(k),U_{e}, P)} = \\ 
		\min\limits_{U_{e}}~ &||Q_{f} x_{e}(N|k)||_{2}+ \sum\limits_{i=0}^{N-1} \left({||Q x_{e}(i|k)||_{2} + ||R u_{e}(i|k)||_{2}} \right)  \\
		\text { s.t. }~ & x_{e}(i+1|k)= A(\rho(i|k))x_{e}(i|k)+B(\rho(i|k)) u_{e}(i|k), \\
		&x_{e}(0|k) = x_{0} - {x}^{\ast}_{0}, ~x_{e}(k) \in \delta\mathbf{X}, \\
		&U_{e} = col\left\{u_{e}(0|k), ~\dots, u_{e}(N-1|k) \right\}, u_{e}(i|k) \in \delta\mathbf{U}, \\
		&P = col\left\{\rho(0|k), \dots, \rho(N-1|k) \right\},	
	\end{aligned}
\end{equation} 
where $J(x_{e}(k),U_{e}, P)$ is the MPC cost function, it is selected as a quadratic function of tracking error and control inputs. $\delta\mathbf{U}$, $\delta\mathbf{X}$ are the set constraints of tracking control inputs such that the actual control and states still fall in the set $\mathbf{U}$, $\mathbf{X}$.

In this section, the state vector with the rear vehicle body coordinate is used during problem formulation. As mentioned in Section~\ref{nl_model}, we can also use the other state representation with the front body coordinate. In this paper, the state vector with front body coordinate is used when driving forward, because the rear vehicle body tends to follow the desired trajectory if the front body tracking error is small. On the contrary, if the state vector with the rear body coordinate is used when driving forward, the tracking error of the front body is not directly considered by the MPC controller. Since the front body position is a function of the rear body position, heading angle, and articulated angle, the tracking errors can also accumulate and cause a large tracking error of the front body, which can even make the system unstable. During application, the state vectors are selected depending on the driving direction of the wheel loader \cite{shi2020planning}. When the wheel loader is moving forward (Step 2 and 4 in Fig.~\ref{fig:loadingcycle}), the front body coordinate needs to be considered in the state vector. Similarly, the rear body coordinate is considered when the wheel loader is reversing (Step 1 and 3 in Fig.~\ref{fig:loadingcycle}). The tracking control system switches from one controller to the other at the loading, unloading, and turning point. This state vector selection method can greatly improve the tracking performance. 

\section{Trajectory Planning and Control System}

\begin{figure}[ht]
\centering
\includegraphics[width = 0.6\linewidth]{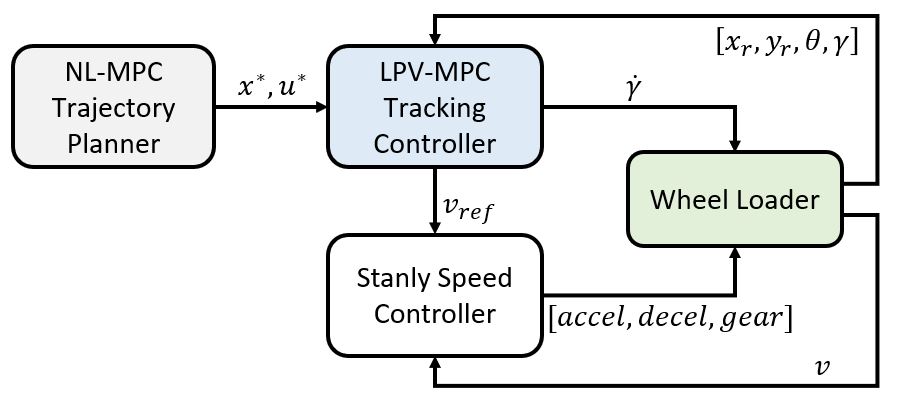}
\caption{Overall architecture of the wheel loader trajectory planning and tracking control system.}
\label{fig:algo_arch}
\end{figure}

Fig.~\ref{fig:algo_arch} is an overview of the whole trajectory planning and control system. The nonlinear MPC trajectory planner sends the desired trajectory points to the LPV-MPC controller who calculates the reference speed, $v_{ref}$, and the articulated angle rate $\dot{\gamma}$ command signals. The reference speed is sent to the Stanly speed controller developed based on \cite{hoffmann2007autonomous}. Since the speed controller is out of the scope of this paper, the details will not be introduced here. 

In the real-world application, the states of position $(x_{f},y_{f})$ and $(x_{r},y_{r})$ are usually measurable by GPS module, the heading angle $\theta$ measured by inertial measurement unit, and the articulated angle can be measured by encoders. Therefore, the deviation of current states and nominal state $x^{\ast}$ can be directly measured by sensors and the scheduling parameters are available online.

\section{Simulation Results and Discussion}\label{simulation}
\subsection{High-fidelity simulation environments}
\begin{figure}[ht]
\centering
\includegraphics[width = 0.6\linewidth]{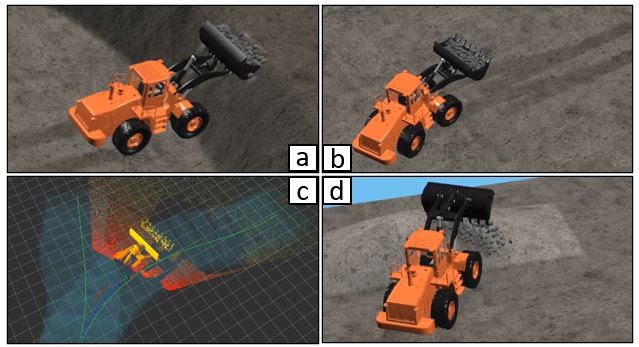}
\caption{A typical loading cycle simulated with AGX Dynamics: (a) Loading material; (b) Transporting material; (c) Following trajectory; (d) Unloading material.}
\label{fig:agx_aes}
\end{figure}
The performance of the proposed planning and control scheme is evaluated in the high-fidelity simulation environment developed by AGX Dynamics \cite{servin2021multiscale} and Matlab/Simulink. As shown in Fig.~\ref{fig:agx_aes}, AGX Dynamics is able to model the wheel loader's tire/soil interaction considering soil deformation, and elasticity, slip, and an-isotropic friction in forward and transverse directions \cite{agxwebsite}. The wheel loader's drive train is simulated by the AGX Drivetrain module with 1D dynamics of engine, clutch, gearbox, and differential. The nonlinear MPC planner and LPV-MPC controller are calculated in Matlab/Simulink and communicate control inputs and states with AGX Dynamics. YALMIP~\cite{lofberg2004yalmip} is used to operate and solve the optimization problem from LPV-MPC, adaptive LTI-MPC, and nonlinear MPC. The solver \textit{Sedumi}~\cite{sturm1999using} is used to solve the semi-definite programming of LTI-MPC and LPV-MPC. The solver \textit{fmincon} is used to solve the nonlinear programming. We perform the simulation on a Dell Precision 7510 with Intel Core i7-6820HQ CPU @2.70GHz x8, and the simulation is conducted in high-fidelity environment with real-time feasibility. The MPC controllers are computed at 5 Hz with a prediction horizon of 10 steps.

\subsection{Simulation results}

\begin{table}[h]
\caption{Nonlinear MPC planner design parameters.}
\label{tab:plan_para}
\begin{center}
\begin{tabular}{c||c}
\hline
 Parameter & Value\\
\hline
$R$ & $diag(1,~1)$\\
$R_d$ & $8 *diag(1,~3)$\\
$\gamma_{max}$ & $0.40~rad$\\
$T_s$ & $0.2~s$\\
$N$ & $100$\\
\hline
\end{tabular}
\end{center}
\end{table}

To compare the performance of the nonlinear MPC, LPV-MPC, and adaptive LTI-MPC controllers, we have conducted the simulation of one whole loading cycle shown in Fig.~\ref{fig:loadingcycle}. The wheel loader first moves from the loading point to the unloading point at the dump truck, and then returns back to the loading point. The trajectory is generated by the nonlinear MPC planner offline with given loading and unloading points. The dump truck and material pile are approximated and considered as rectangular obstacles by the nonlinear MPC planner. The main parameters are shown in Table~\ref{tab:plan_para}.

\begin{table}[h]
\caption{MPC controller design parameters.}
\label{tab:ctr_para}
\begin{center}
\begin{tabular}{c||c}
\hline
 Parameter & Value\\
\hline
$R$ & $diag(0.1,~0.5)$\\
$Q$ & $8 *diag(4,~4,~3,~2)$\\
$Q_f$ & $10*Q$\\
$T_s$ & $0.2~s$\\
$N$ & $10$\\
\hline
\end{tabular}
\end{center}
\end{table}

\subsubsection{Tracking performance}
\begin{figure*}[ht]
\centering
\includegraphics[clip, width = 0.95\linewidth]{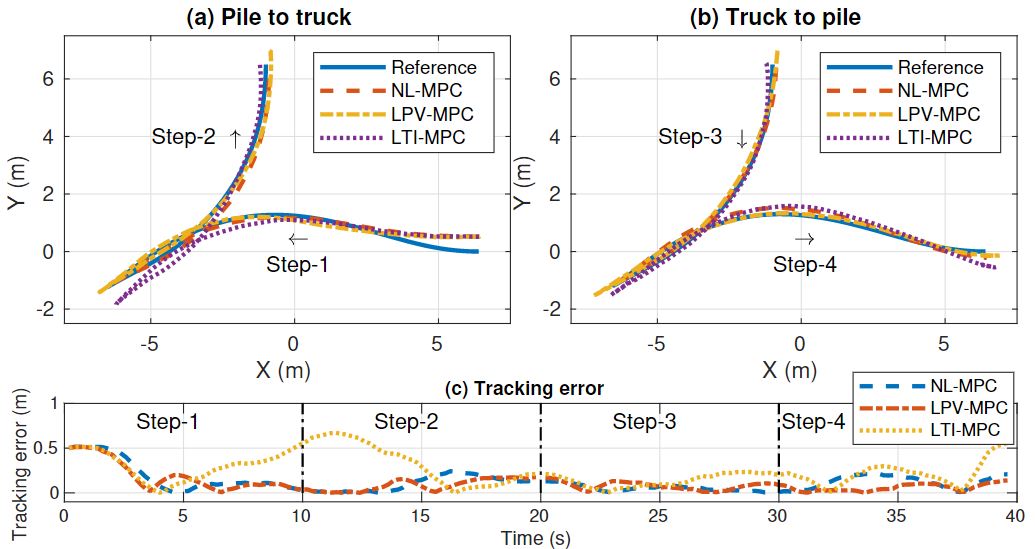}
\caption{Wheel loader position tracking performance of NL-MPC, LPV-MPC, and LTI-MPC controllers.}
\label{fig:xy_track}
\end{figure*}

Once the offline trajectory is obtained with position, heading, and articulated angle information, the wheel loader can be controlled by the three MPC controllers to track the trajectory. To make a fair comparison, all the controllers share the same design parameters, as shown in Table~\ref{tab:ctr_para}. To demonstrate the stability of the controller, we intentionally added 0.5m lateral error at the beginning of the simulation. Fig.~\ref{fig:xy_track} shows the rear vehicle body coordinates $\left(x_r, y_r\right)$ provided by the planned trajectory along with the vehicle response of the three different MPC approaches. Subplot~(a) shows the actual trajectories starting from the same initial state, where the wheel loader locates around the pile point, both the heading angle and the articulated angle are 0. Subplot~(b) shows the actual returning trajectory from the truck to the pile. Subplot~(c) shows the absolute trajectory tracking error. The tracking error is calculated by the distance of the wheel loader's current position to the nearest point on the trajectory.

The nonlinear MPC and LPV-MPC are able to accurately follow the desired trajectory during the whole loading cycle in the high-fidelity environment and have similar performance. The LTI-MPC has the worst tracking performance since it fails to accurately capture the kinematics of the wheel loader especially when the vehicle heading and articulated angle are changing quickly. The tracking error leads the LTI models to deviate from the actual model along the trajectory, which deteriorates the tracking performance. On the contrary, the LPV model includes nominal nonlinear dynamics in the prediction and produces more accurate tracking than the adaptive LTI-MPC. Table~\ref{tab:track_err} lists the mean of absolute tracking errors of the three controllers during the whole loading cycle.

\begin{table}[h]
\caption{Tracking Error Comparison}
\label{tab:track_err}
\begin{center}
\begin{tabular}{c||ccc}
\hline
 & NL-MPC & LPV-MPC & LTI-MPC\\
\hline
\begin{tabular}{@{}c@{}}Mean absolute\\tracking error (m)\end{tabular} & 0.103 & 0.120 & 0.246\\
\hline
\end{tabular}
\end{center}
\end{table}

\begin{figure}[ht]
\centering
\includegraphics[clip, width = 0.7\linewidth]{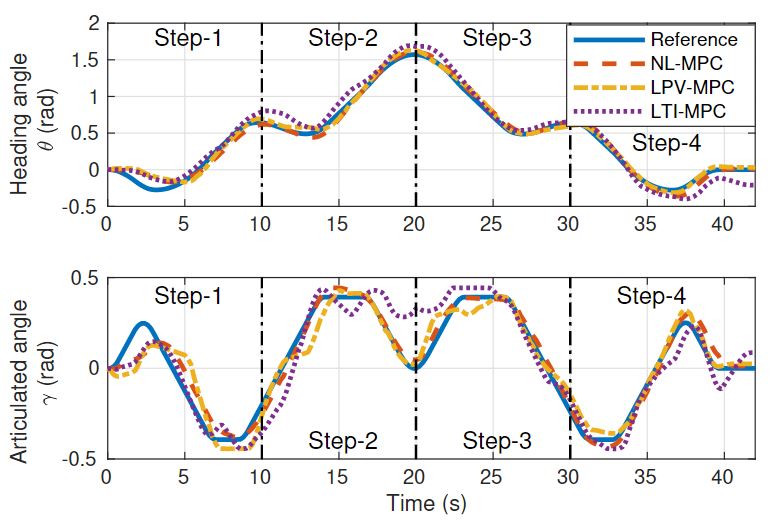}
\caption{Wheel loader heading angle and articulated angle tracking performance of NL-MPC, LPV-MPC, and LTI-MPC controllers.}
\label{fig:angle_track}
\end{figure}

For a wheel loader, the coordinate of the front or rear vehicle body cannot fully define the pose of the vehicle. Heading and articulated angles also need to be considered while evaluating the tracking performance. When the wheel loader is away from the loading or unloading points, the heading and articulated angles are less important as long as the vehicle's position can track the desired path without collision with obstacles. However, tracking the planned heading and articulated angles helps controllers to manipulate the vehicle's position closer to the desired path. When the wheel loader is at the loading or unloading point, it is expected that the heading and articulated angles can match the desired values closely, since these two angles greatly affect the wheel loader's loading and unloading maneuvers. Fig.~\ref{fig:angle_track} illustrates the desired and actual heading and articulated angles during simulation. The nonlinear MPC and LPV-MPC generally have better tracking performance than the LTI-MPC, especially at the loading and unloading points (at time = $20s$ and $40s$). 

\begin{figure}[ht]
\centering
\includegraphics[clip, width = 0.7\linewidth]{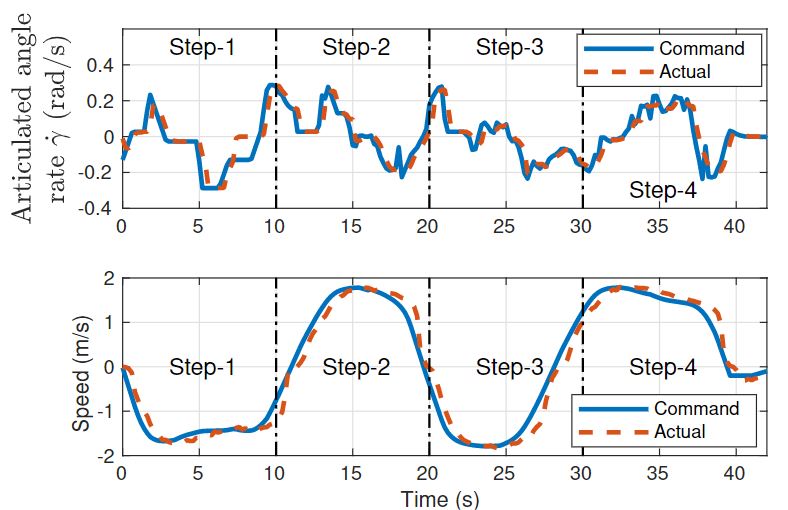}
\caption{Control actions of LPV-MPC controller.}
\label{fig:control_cmd}
\end{figure}

Fig.~\ref{fig:control_cmd} plots the control actions calculated by the LPV-MPC controller and the corresponding response measured from the wheel loader AGX model. The speed response has a larger delay and tracking error than the articulated angle response, since we are not considering the speed as the state in the vehicle model. The performance of the speed controller is out of the scope of this paper, so will not be discussed here. However, it needs to be noted that the tracking error of the speed controller acts as one of the disturbances of the whole system and greatly affects the performance of MPC controllers used for trajectory tracking.

\subsubsection{Computational burden}
\begin{figure}[ht]
\centering
\includegraphics[clip, width = 0.7\linewidth]{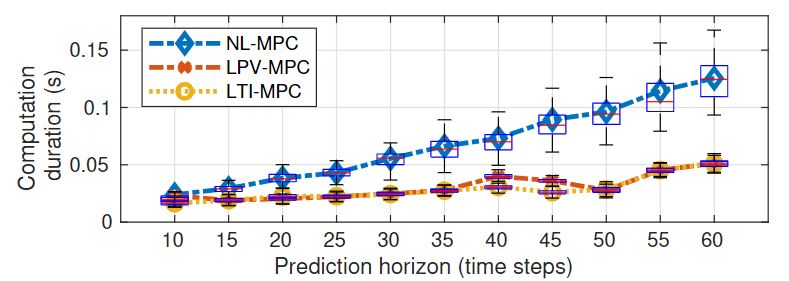}
\caption{Computation burden comparison.}
\label{fig:computationalBurden}
\end{figure}

With the tracking performance comparison in mind, we further compare the computational burdens of three MPC strategies. The computational burdens are plotted in Fig.~\ref{fig:computationalBurden} with average computational duration and variances. It is obvious that the computational duration of nonlinear MPC grows much faster than the adaptive LTI-MPC and LPV-MPC with increasing prediction horizons. However, the LPV-MPC has almost the same computational duration as LTI-MPC. This result indicates that the LPV-MPC renders similar computational complexity with LTI-MPC, and much less complexity than nonlinear MPC. The reason is that both the LPV-MPC and LTI-MPC formulate quadratic programs in each step, but the nonlinear MPC has to solve the nonlinear programming. 

Combining the results of tracking performance and computational burden, the LPV-MPC strategy produces close tracking performance with nonlinear MPC but greatly reduces the computational complexity. Comparing with the adaptive LTI-MPC strategy, the LPV-MPC has a similar computational burden but produces much more accurate tracking performance. It needs to be noted that LPV-MPC cannot be easily used for planning, because the LPV-MPC algorithm requires the knowledge of the scheduling parameters within the prediction horizon which is not available before planning.

\section{Conclusions and Future Work}
This paper presents a systematic framework of offline high-performance planning and online tracking by LPV-MPC strategy. With given initial and target final states, the trajectory planning is conducted by nonlinear MPC with nonlinear models. The proposed LPV-MPC is used to online track the planned trajectory. In the high-fidelity simulation, the LPV-MPC is demonstrated to perform much better than LTI-MPC in tracking performance and faster computation than nonlinear MPC while maintaining good tracking performance. 

In the high-fidelity simulation, the wheel loader is subject to disturbance from frictions between wheels and uneven, uncertain ground. However, the robustness against these disturbances and model uncertainty by the LPV-MPC is not analyzed. The future work is to develop tools for robustness analysis and to develop a robust LPV-MPC strategy to reject external disturbance and achieve guaranteed robustness against model uncertainty. The current implementation does not consider the full dynamics of the wheel loader, so another future work is to design systems considering the full dynamics of the vehicle including the case when the dynamics are changing due to the load being carried by the wheel loader.

%Bibliography
\bibliographystyle{unsrt}  
\bibliography{reference}

\end{document}